\documentclass[a4paper,aps,manuscript]{revtex4}
\usepackage[T1]{fontenc}
\usepackage[latin9]{inputenc}
\usepackage{amsmath}
\usepackage{amssymb}
\usepackage{graphicx}
\usepackage{esint}

\makeatletter

\@ifundefined{textcolor}{}
{%
 \definecolor{BLACK}{gray}{0}
 \definecolor{WHITE}{gray}{1}
 \definecolor{RED}{rgb}{1,0,0}
 \definecolor{GREEN}{rgb}{0,1,0}
 \definecolor{BLUE}{rgb}{0,0,1}
 \definecolor{CYAN}{cmyk}{1,0,0,0}
 \definecolor{MAGENTA}{cmyk}{0,1,0,0}
 \definecolor{YELLOW}{cmyk}{0,0,1,0}
}

\makeatother

\begin{document}

\title{Fluctuation driven height reduction of crosslinked polymer brushes.
A Monte Carlo Study.}

\author{M. Lang$^{1*}$, M. Hoffmann$^{1,2,3}$, M. Werner$^{1,2}$, R. Dockhorn$^{1,2}$,
and J.-U. Sommer$^{1,2}$}

\affiliation{$^{1}$\emph{Leibniz-Institute of Polymer Research Dresden, Hohe
Straße 6, 01069 Dresden, Germany} $^{2}$\emph{Institute of Theoretical
Physics, Technische Universität Dresden, Zellescher Weg 17, 01062
Dresden, Germany} $^{3}$\emph{Current address: BioQuant \& Institute
of Theoretical Physics, Heidelberg University, 69117 Heidelberg, Germany}}
\begin{abstract}
We study the changes in the conformations of brushes upon the addition
of crosslinks between the chains using the bond fluctuation model.
The Flory-Rehner model applied to uni-axially swollen networks predicts
a collapse for large degrees of crosslinking $q$ proportional to
$q^{-1/3}$ in disagreement with our simulation data. We show that
the height reduction of the brushes is driven by monomer fluctuations
in direction perpendicular to the grafting plane and not due to network
elasticity. We observe that the impact of crosslinking is different
for reactions between monomers of the same or on different chains.
If the length reduction of the effective chain length due to both
types of reactions is accounted for in a function $\beta(q)$, the
height of the brush can be derived from a Flory approach for the equilibrium
brush height leading to $H(q)\approx H_{b}\beta(q)^{1/3}$, whereby
$H_{b}$ denotes the height of the non-crosslinked brush. 
\end{abstract}
\maketitle

\section{Introduction}

If polymer chains are grafted to a surface densely enough, the chains
stretch out due to excluded volume interactions. Alexander and de
Gennes developed the first theoretical models for polymer brushes
based on mean-field and scaling concepts~\cite{alexander:scaling,degennes:scaling,Halperin1992}.
Later, Semenov~\cite{Semenov} introduced a parabolic molecular field
to describe the conformations of blocks copolymers in the strong segregation
limit. Milner \textit{et al.}~\cite{milner:scf:epl} and Skvortsov
\emph{et al}.~\cite{skvortsov:brushes1} obtained a similar refined
description of the monomer profile inside the brush using self-consistent
field approaches. In the scaling approach, the chains are considered
as a stretched array of correlation blobs at a grafting density $\sigma$
above the overlap grafting density $\sigma^{*}\propto R^{-2}\propto b{}^{-2}N{}^{-2\nu}$.
Here, $b$ is the root mean square bond length, $N$ the degree of
polymerization of the chains, $\nu\approx0.588$ the exponent for
chains in athermal solvents ~\cite{Rubinstein}, and $R$ is the
size of a free coil. The correlation length $\xi$ is determined by
the grafting density $\xi\approx\sigma^{-1/2}$. Each correlation
volume contains 
\begin{equation}
g\approx(\xi/b)^{1/\nu}\approx\sigma^{-1/(2\nu)}b^{-1/\nu}\label{eq-g}
\end{equation}
monomers and the equilibrium brush height is given by 
\begin{equation}
H_{b}\approx\xi N/g\approx N\sigma^{(1-\nu)/(2\nu)}b^{1/\nu}.\label{eq-height}
\end{equation}

Several groups tested these predictions and explored detailed static
and dynamic properties of polymer brushes by applying Monte Carlo
or Molecular Dynamics simulations~\cite{binder:structure,binder:mcs,binder:theta,murat:structure,murat:solvent,grest:brushes,Merlitz:Brush,Merlitz3,Merlitz4,Merlitz5}.
On the experimental side, large progress has been made over the years~\cite{advincula}
with recent accomplishments allowing very high grafting densities
to be examined~\cite{tsujii:brushes}. Since the polymers in the
brush control the net interaction of the surface with its environment,
this gave rise to a large variety of different applications. Examples
are drug delivery~\cite{torchilin:drug}, colloid stabilization~\cite{Napper,Russel},
reduction of friction~\cite{klein:brush1,klein:brush2,Moro_NATMAT04},
increasing the bio-compatibility of medical implants~\cite{Zdyrko:Brushimplant},
and switchable amphiphilic surfaces~\cite{Uhlmann:Brush2,uhlmann}.

For these applications, polymer brushes are often situated on surfaces
that are exposed to a harsh environment. This can cause a destruction
of the brush and chains may be broken or torn away from the substrate.
To reduce degradation, crosslinking the chains is one obvious alternative.
In a crosslinked brush, chains form a connected network, which prevents
individual chains from leaving the surface layer even if their grafting
points are broken. Therefore, crosslinking was used previously for
stabilization \cite{guojun,liu}, while it was also used to freeze
a certain state of switchable brushes~\cite{Uhlmann:Brush2}.

In our preceding publication \cite{Hoffmann}, we explored the universality
of the network structure and the linking statistics parallel to the
grafting plane. It was found that a crosslinked brush can be mapped
onto a two dimensional percolation problem, whereby the overlap of
the chains controls the percolation problem. In the present work,
we investigate the changes of the static behavior of polymer brushes
upon crosslinking for systems far away from the gel point. In particular,
we focus on the experimentally relevant question of the height reduction
of the brush due to crosslinking. For a theoretical description of
this problem, we derive the uni-axial swelling behavior of a crosslinked
grafted layer using two different approaches: the Flory-Rehner Model
in section \ref{sec:The-Equilibrium-Height} and a new approach considering
monomer fluctuations in direction perpendicular to the grafting plane
in section \ref{Crosslinked}. This latter approach predicts that
inserting elastically active crosslinks into a brush does not collapse
the brush (up to logarithmic corrections as function of the degree
of polymerization $q$). This result is in clear contrast to the Flory-Rehner
prediction, which leads to a collapse $\propto q^{-1/3}$. In order
to directly assess the change of brush height upon crosslinking, we
perform a parallel analysis of mono-disperse crosslinked and non-crosslinked
brushes with the same degrees of polymerization $N$ and grafting
densities $\sigma$ and compare with the predictions of both models.
The details of the simulations are described briefly in the following
section.

\section{Simulations\label{Methods}}

We use the Bond-Fluctuation-Model (BFM) as introduced in 1988 by Carmesin
and Kremer~\cite{carmesin:bfm} and extended to three dimensions
by Deutsch and Binder~\cite{deutsch:interdiffusion}. In this model,
each monomer is represented by a cube of eight lattice positions on
a regular cubic lattice. Excluded volume is modeled by not allowing
monomers to overlap. The monomers of one chain are connected by a
bond vector out of a predefined vector set. The combination of bond
vector set and excluded volume of monomers ensures cut avoidance of
strands without an explicit test of the local topology~\cite{deutsch:interdiffusion}.
Solvent is treated implicitly in the athermal limit and hydrodynamics
is neglected. The BFM is well suited to capture static and dynamic
properties of polymers melts~\cite{deutsch:interdiffusion}, solutions~\cite{paul1991crossover},
and networks~\cite{Sommer:Bfm1,Sommer:Bfm2,sommer_MM02,lang_PRL10,lang_MM05}.

The present work analyzes the same brushes as our previous work and
we refer the reader for details on crosslinking and equilibration
to Ref. \cite{Hoffmann}. In brief, we created brushes with degrees
of polymerization $N=16$, 32, and 64 and grafting densities $\sigma=1/256$,
1/64, 1/25, 1/16, 1/9, 4/25, and 1/4. Note that the grafting density
$\sigma$ is normalized to the maximum possible grafting density in
the BFM model, which is 0.25 in square lattice units. System sizes
varied between approximately 1500 and 10000 grafting points arranged
regularly on the $xy$-plane. Periodic boundary conditions were applied
in $xy$-direction and reflecting boundary conditions in the $z$-direction.
After equilibration, crosslinking occurs upon collision of two monomers.
We only allow one additional crosslink per monomer. Crosslinking is
stopped after a defined number of crosslinks in the sample is established.
Note that we varied the crosslinking rate from 1 to 0.001 and did
not observe a significant modification of the results. Thus, the networks
of the present paper were reacted with the maximum possible reaction
rate 1. After crosslinking, brushes are relaxed until no further changes
in the brush profile and the average monomer fluctuations were recognized.

\section{The Equilibrium Height of a Uni-axially Swollen Network\label{sec:The-Equilibrium-Height}}

In this section, we assume that a crosslinked brush can be modeled
analogous to a uni-axially swollen network. The chains of a polymer
brush are concentrated above their overlap concentration. Therefore,
introducing additional crosslinks can shrink the brush height only,
if the network modulus becomes larger than the osmotic pressure of
the grafted chains. On a scaling level, this is equivalent to the
condition, that there must be of order one crosslink per blob of the
brush. Let us introduce the average number $q$ of crosslinks per
chain. We further assume a perfect network structure for simplification.
Since any crosslink connects two strands, the average strand length
$n$ of a network chain is given by 
\begin{equation}
n=N/(2q+1).\label{eq:strand}
\end{equation}
Therefore, if this strand length $n$ is larger than the number $g$
of monomers per blob, $n>g$, we expect that the height of the brush
remains mainly unmodified. Otherwise, for $n<g$, it is expected that
the modulus of the network controls the height of the brush.

For this latter case, we use the Flory-Rehner (FR) approach \cite{FloryRehner}
for network swelling to estimate the shrinking of the brush as function
of crosslinking. Here, the mean field estimate for the osmotic pressure
\begin{equation}
\Pi(\phi)=\frac{k_{B}T}{b^{3}}\left[\frac{\phi^{2}}{2}\left(1-2\chi\right)+\frac{\phi^{3}}{3}+...\right]\label{eq:Pi-1}
\end{equation}
is balanced by the Gaussian elasticity 
\begin{equation}
G(\phi)=\frac{\nu_{2}k_{B}T}{2}\frac{(\lambda R)}{b^{2}N}^{2}\approx\frac{\phi}{n}\frac{k_{B}T}{2b^{3}}\frac{(\lambda R)}{b^{2}N}^{2}\label{eq:G-1}
\end{equation}
of the elastic strands as estimated by the phantom network model.
$\nu_{2}$ is the number density of elastic strands, $\chi$ is the
Flory interaction parameter, $\phi$ is the volume fraction of polymer,
$\lambda$ is the deformation ratio of the elastic strands, $k_{B}$
is the Boltzmann constant, and we used $b^{3}$ as estimate for the
monomeric volume. In order to apply this approach to the equilibrium
height of a crosslinked brush we have to consider the particular geometry
of the problem and the crosslinking of the chains.

In the Flory-Rehner approach \cite{FloryRehner}, one considers the
dry state (no solvent) as reference state for crosslinking. Swelling
from the dry state is equivalent to a uni-axial deformation at constant
cross-section of the network. This is expressed by introducing deformation
ratios $\lambda_{x}=\lambda_{y}=1$ parallel to the grafting plane
and 
\begin{equation}
\lambda_{z}=H/H_{dry}=1/\phi\label{eq:lambda}
\end{equation}
in direction of swelling. Assuming affine deformation of the strands,
the change in elastic free energy per strand can be written as 
\begin{equation}
\Delta f_{el}=-T\Delta S_{el}=\frac{k_{B}T}{2}\left(\lambda_{z}^{2}-1\right)=\frac{k_{B}T}{2}\left(\frac{1-\phi^{2}}{\phi^{2}}\right).\label{eq:DFe2-1}
\end{equation}
Therefore, the density dependence of modulus can be written as 
\begin{equation}
G(\phi)\approx\frac{\phi}{n}\frac{k_{B}T}{2b^{3}}\left(\frac{1-\phi^{2}}{\phi^{2}}\right).\label{eq:G-1-1}
\end{equation}
Equilibrium brush height is achieved %
\footnote{Equilibrium is really attained by minimizing the free energy $\partial F/\partial V=0$.
But since both leading terms for the free energy are power laws in
concentration, the present simplified discussion is correct up to
a numerical constant \citep{Rubinstein}.%
} for $\Pi(\phi)=G(\phi)$, which leads to 
\begin{equation}
\phi^{2}\left(1-2\chi\right)+\frac{2\phi^{3}}{3}=\frac{(1-\phi^{2})}{\phi n}.\label{eq:eqFR-1}
\end{equation}
For large degrees of swelling, $\phi\ll1$, we drop the third virial
and the $\phi^{2}$-term of the modulus to approximate 
\begin{equation}
\phi\approx\left(\frac{1}{n(1-2\chi)}\right)^{1/3}.\label{eq:eq_phi-1}
\end{equation}
Thus, the Flory-Rehner model predicts for an uni-axially swollen network
a continuous shrinking as function of crosslinking: 
\begin{equation}
H\propto n^{1/3}\propto\left(N/(2q+1)\right)^{1/3}.\label{eq:H_FR-1}
\end{equation}

As mentioned above, height reduction due to network elasticity becomes
only effective, if $g>n$. Therefore, we have for $g\lesssim n$ the
height of the non-crosslinked brush $H_{b}$ as given in equation
(\ref{eq-height}) and for $g>n$ we expect that 
\begin{equation}
H\approx H_{b}\left(\frac{n}{g}\right)^{1/3}.\label{eq:shrink}
\end{equation}
Note that a scaling model for the uni-axial swelling of a network
leads to similar results.

We use the z-component of the center of mass of all monomers $z_{{\rm CM}}$
as indicator for the brush height $H\propto z_{{\rm CM}}$. It is
computed by 
\begin{equation}
z_{{\rm CM}}=\frac{1}{NM}\sum_{m=1}^{M}\sum_{i=1}^{N}z_{m,i},\label{eq-brush-height-simulated}
\end{equation}
where $M$ is the number of chains in the system and $z_{m,i}$ is
the z-position of the $i$-th monomer (as counted from the grafting
point) of chain $m$. The number of monomers per blob, $g,$ is estimated%
\footnote{Note that for simplification, we use the geometrical $g=(\xi/b)^{1/\nu}=[2/(b\sigma^{1/2})]^{1/\nu}$
to analyze the simulation data, whereby the factor 2 results from
the normalization of $\sigma$ per lattice unit.%
} using equation (\ref{eq-g}).

As reference for the brush size we use the height of the non-crosslinked
brush, $H_{b}$ as determined from the simulation data. In brief,
for $H_{b}$ we found good agreement with the data of previous works~\cite{binder:scaling,binder:structure,seidel,Chakrabarti:Brush,Merlitz:Brush}:
the height of the non-crosslinked brushes follows the prediction of
equation (\ref{eq-height}) except of small non-Gaussian corrections
due to the overstretching of the chains at large $\sigma$ as discussed
previously~\cite{Shim,Merlitz:Brush,Biesheuvel}. Similar observations
were made for the parabolic density profile as predicted by self-consistent
field (SCF) approaches~\cite{Semenov,milner:scf:epl,milner:scf:mm,milner:brushes,skvortsov:brushes1,skvortsov:brushes2}
including small deviations to these predictions as discussed in previous
work \cite{Shim,Merlitz:Brush,Biesheuvel}.

\begin{figure}
\includegraphics[angle=270,width=0.8\columnwidth]{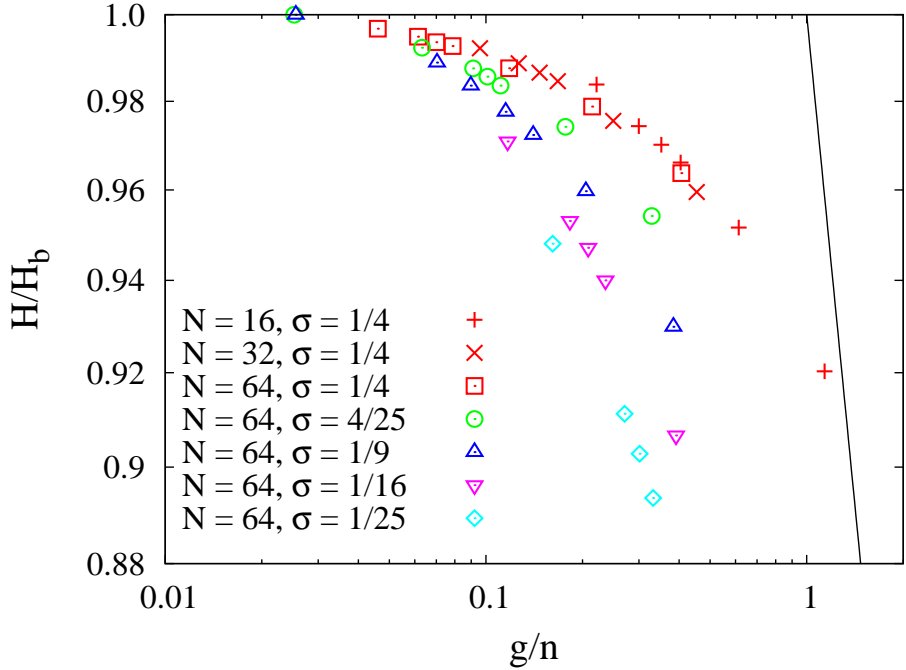}

\caption{\label{fig:Height-reduction-of}(Color online all Figures) Height
reduction of the brush as function of crosslinking. The line indicates
the prediction of equation (\ref{eq:shrink}) for comparison.}
\end{figure}

The height reduction of the brushes is analyzed as function of the
degree of crosslinking $q$ using the ratio $g/n$ in Figure \ref{fig:Height-reduction-of}.
For a better comparison, we determine $q$ only from ``real crosslinks''
that link monomers of different chains and ignore all links between
the monomers of the same chain (``self-links''), which do not contribute
to elasticity. We observe that most crosslinked brushes remain in
the brush dominated regime $n>g$ and no data is clearly in the network
regime even though we linked a large fraction of monomers (in some
cases up to 1/3 of all monomers). The most important observation is
that the data do not collapse as a function of $g/n$. This suggests
that there might be a mechanism different to network elasticity that
drives the collapse of the brush upon crosslinking. This mechanism
is discussed in the following section and we consider it as the main
source for height reduction when crosslinking a brush in the swollen
state.

\section{Fluctuation Driven Height Reduction of a Brush Crosslinked in the
Swollen State\label{Crosslinked}}

The idea of our approach is that the fluctuations of the monomers
in $z$-direction determine an average length reduction of the effective
strands of the brush, since monomers with different index $i$ or
$j$ as counted from the grafting point are connected as shown in
Figure \ref{fig:Grafik} a) and b). We aim to show, that this length
reduction determines the height reduction of the brushes as function
of the degree of crosslinking $q$.

Since the contacts along the same chain and among different chains
follow different statistics, we distinguish throughout this section
between links that connect monomers of the same chain (``self-links'')
from links that connect monomers of different chains (``crosslinks''),
see Figure \ref{fig:Grafik} a) and b) respectively. In order to demonstrate
that the above proposal can be used to understand the height reduction
of swollen brushes upon crosslinking, we first analyze the length
reduction of the effective chains inside the brush and compare with
the simulation data. In a second step, the result for the length reduction
is used to compute the height reduction by constructing an appropriate
Flory estimate for the brush height.

\subsection{Length reduction of minimal and effective chains inside the brush.}

Let us first introduce $N_{min}$ as the minimum number of bonds along
the network structure between a chain end and the grafting plane.
We also introduce the effective chain length $N_{x}$ as the equivalent
chain length to describe the modified elasticity of the chain as part
of a brush after the crosslinking reactions. Since $N_{x}$ is difficult
to determine from the network data while the determination of $N_{min}$
is rather trivial, we present here all simulation data as function
of $N_{min}$ and use the mean field relations between the average
$N_{min}$ and $N_{x}$ as derived below to analyze the height reduction
of the brushes in the following subsection.

\begin{figure}
\includegraphics[width=0.8\columnwidth]{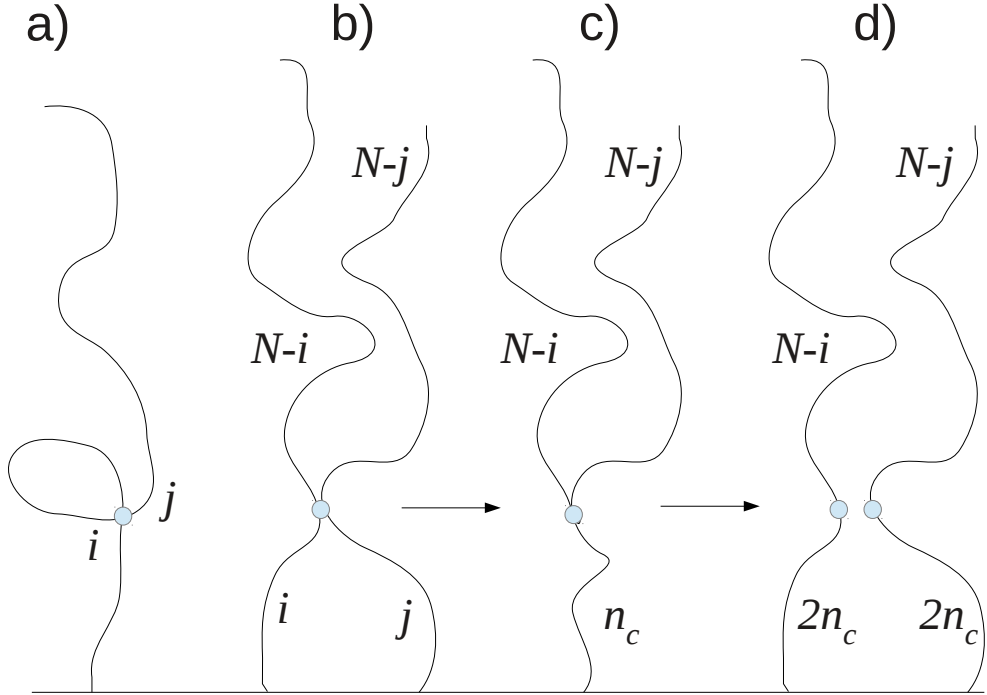} \caption{\label{fig:Grafik}Reduction of effective chain length upon crosslinking.
Monomer $i$ and $j$ form a ``self-link'' along chain a). The monomers
$i$ and $j$ of the two chains at b) are ``crosslinked''. c) and
d) describe the two steps for computing the effective elastic chain
lengths $N_{x}$ as discussed in the text.}
\end{figure}

\textbf{Self-links}\textbf{\emph{ }}lead to dangling loops as shown
in Figure \ref{fig:Grafik} a). The length distribution of these loops
is dominated by links to the nearest neighbors similar as for linking
chains in melts~\cite{lang_MM05}. This is because the return probability
decays quicker than $|i-j|^{-1}$ for the self-avoiding walk inside
the blob $|i-j|<g$ and for the stretched sections with $|i-j|>g$.
The largest contribution to length reduction is thus dominated by
the shortest loops. For the intermediate grafting densities of the
present study we conclude, therefore, that the length reduction due
to self-links is approximately independent of $\sigma$. Since each
self-link roughly leads to the same total change in $N_{min}$, we
expect that the total length reduction $N-\langle N_{min}(q)\rangle$
due to self-links is proportional to $q$ in first approximation,
thus 
\begin{equation}
\langle N_{min}(q)\rangle/N=1-\gamma q/N.\label{eq:self}
\end{equation}
The brackets denote here the sample average and $\gamma$ is average
number of monomers per dangling loop.

The monomer fluctuations in $z$-direction control the average length
reduction $N-\langle N_{min}(q)\rangle$ due to \textbf{crosslinking}
different chains. Let us assume a parabolic density profile for the
non-crosslinked brush. Then, the fluctuations of monomer $i$ in $z$-direction
are roughly $\propto Hi/N$ and thus, depend on the grafting density
$\sigma$. In conclusion, the average difference $\left\langle |i-j|\right\rangle $
of any two monomers in contact inside an non-crosslinked brush equals
the average fluctuation of a strand of $\left\langle i+j\right\rangle $
monomers in $z$-direction and thus, $\left\langle |i-j|\right\rangle =cN$
using a parameter $c$ to describe the average fluctuations in $z$-direction
of a monomer inside the brush.

Let us now consider by $i$ and $j$ the \emph{minimum number} of
bonds between these monomers and the grafting plane. Each crosslink
equalizes the minimal distance along the strands to the grafting point
at the connected monomers. Each equalization reduces the possible
differences $\left\langle |i-j|\right\rangle $ of the neighboring
connected chains in the same manner as the total average monomer fluctuations
in $z$-direction become proportional to the average length of the
elastic network strands $n$. Therefore, we conclude that the \emph{rate
of length reduction} during crosslinking is 
\begin{equation}
\propto\left\langle |i-j|\right\rangle \propto cn\propto cN/(2q+1).\label{eq:diff}
\end{equation}
Since the total length reduction is the integral over this rate up
to crosslinking degree $q$, the average length reduction $N-\left\langle N_{min}(q)\right\rangle $
is estimated as 
\begin{equation}
N-\left\langle N_{min}(q)\right\rangle \approx\int_{0}^{q}\frac{cN}{(2q'+1)}\mbox{d}q'\approx cN\ln(2q+1)/2.\label{eq:fluct}
\end{equation}
The relative length reduction $\langle N_{min}(q)\rangle/N$ is therefore
a non-linear function of the degree of crosslinking 
\begin{equation}
\frac{\langle N_{min}(q)\rangle}{N}\approx1-c\ln(2q+1)/2.\label{eq:cross}
\end{equation}

For testing equation (\ref{eq:self}) and (\ref{eq:cross}) we separately
analyze the length reduction $\langle N_{min}(q)\rangle/N$ due to
self-links and crosslinks. To this end, we virtually remove either
all self-links or crosslinks from the sample and determine $N_{min}$
from the remaining bonds. Also $q$ is computed from the remaining
bonds for each particular sample.

\begin{figure}
\includegraphics[angle=270,width=0.8\columnwidth]{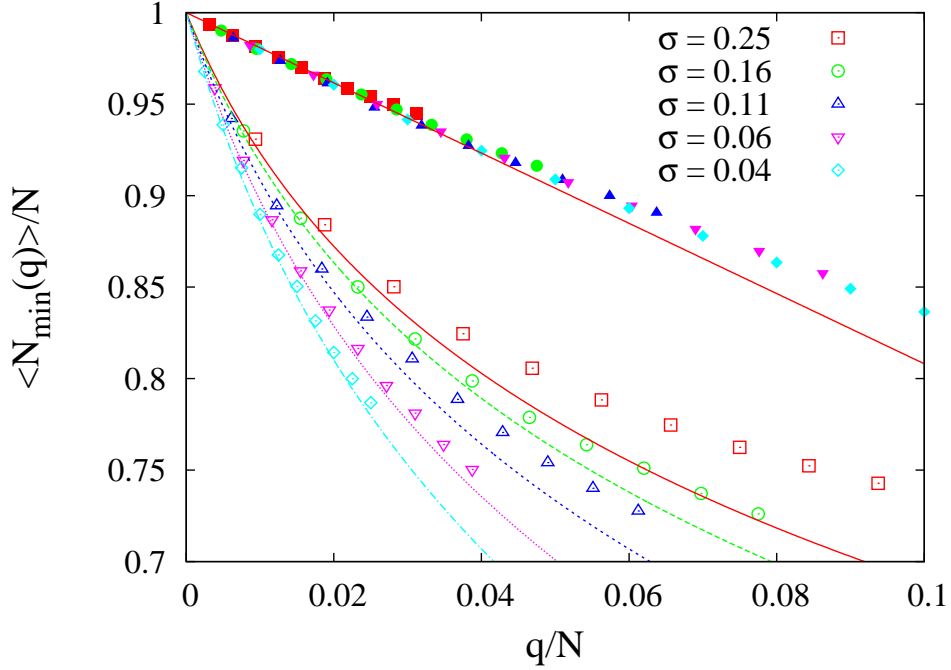} \caption{\label{fig:l_reduction}Length reduction $\langle N_{min}(q)\rangle/N$
of chains $N=64$ due to self-links (full symbols) and crosslinks
(hollow symbols) as function of $q/N$.}
\end{figure}

Figure \ref{fig:l_reduction} shows that the self-link contribution
is essentially independent of $\sigma$ and proportional to $q$ as
proposed above. The slight deviation between data and linear fit is
due to the increased tension along the chains upon crosslinking and
the resulting reduced average size of a dangling loops for high degrees
of crosslinking. A fit of the data indicates that at the beginning
of the reactions, in average about $\gamma\approx1.9$ monomers are
part of a dangling loop, while this drops to about $\gamma\approx1.5$
monomers for large $q$.

We determined relative distance $\left\langle |i-j|\right\rangle /N$
of monomers in contact in the non-crosslinked brush and use these
results as parameter $c$. We obtain $c\approx0.156,$ $0.168,$ $0.187,$
$0.210,$ and $0.233$ for the series of samples from $\sigma=1/4$
to $\sigma=1/25$. Equation (\ref{eq:cross}) is plotted using these
$c$ in Figure \ref{fig:l_reduction} \emph{without} explicitly fitting
the data. The qualitative agreement between the data and our simple
model is remarkable. Note also the qualitative difference between
self- and crosslinks as predicted by our model. For high degrees of
crosslinking we observe the same corrections as for the self-loop
data, which we attribute again to the increased tension along the
chains because of crosslinking. Altogether, the above results are
a sound basis to estimate the effective length of the elastic brush
chains $N_{x}$ after crosslinking.

Obviously $N_{x}=N_{min}$, if only self-links are present, since
the monomers inside a dangling loop do not contribute to elasticity.
For the crosslinks, the parts of chains at Figure 2 b) that lie between
grafting plane and monomer $i$ or $j$ respectively can be described~\cite{Rubinstein}
by a combined chain of length $i+j$ that is grafted at both ends.
These chains act in parallel on the crosslink like two independent
confining potentials of strength $\propto1/i$ and $\propto1/j$.
The fluctuations of the crosslink are thus restricted by a confining
potential $\propto(1/i+1/j)$, which is modeled by an elastic chain
of 
\begin{equation}
n_{c}=ij/(i+j)\label{eq:nc-1}
\end{equation}
segments that connects the grafting plane with the crosslink, see
Figure 2 c). In order to project this result back into the brush problem
we have to equally split this elastic chain $n_{c}$ at \emph{constant
total elasticity} of the system. If two chains of $2n_{c}$ monomers
are connected, the combined chain at the connection point equals an
elastic chain of $n_{c}$ monomers, because of $1/n_{c}=1/(2n_{c})+1/(2n_{c})$.
Therefore, the effective chain lengths after splitting as shown in
Figure \ref{fig:Grafik} d) become

\begin{equation}
N_{x,i}=N-i+2n_{c}\label{eq:neff1}
\end{equation}
\begin{equation}
N_{x,j}=N-j+2n_{c}.\label{eq:neff2}
\end{equation}
This leads to a net change in the total effective chain length of
\begin{equation}
\Delta N_{x}=2N-N_{x,i}-N_{x,j}=\frac{\left(i-j\right)^{2}}{i+j}\label{eq:DeltaNeff}
\end{equation}
per additional crosslink.

There is an average $\left\langle i+j\right\rangle =N$ and an average
$\left\langle |i-j|\right\rangle =cN$ at the beginning of the reaction.
Similarly, $\left\langle i+j\right\rangle =n$ and $\left\langle |i-j|\right\rangle \propto cn\propto cN/(2q+1)$
at later stages of the reaction. Therefore, we can approximate for
the average change in the effective chain length 
\[
\left\langle \Delta N_{x}(q)\right\rangle =\int_{0}^{q}\left\langle \Delta N_{x}(q')\right\rangle \mbox{d}q'\approx\int_{0}^{q}\left\langle \frac{c^{2}N^{2}(2q'+1)}{N(2q'+1)^{2}}\right\rangle \mbox{d}q'
\]
\begin{equation}
\approx c^{2}N\ln(2q+1)/2,\label{eq:NNeq}
\end{equation}
which leads to a relative length reduction of the effective chain
length due to crosslinks 
\begin{equation}
\frac{\left\langle \Delta N_{x}(q)\right\rangle }{N}\approx1-c^{2}\ln(2q+1)/2.\label{eq:DeltaNrel}
\end{equation}

In our previous work~\cite{Hoffmann} we showed that the number of
self-links decays roughly as $\sigma^{-3/2}$ at the gel point, while
the number of crosslinks remains constant. Throughout the reactions
we do not observe a significant shift from self-links to crosslinks
or vice versa. Therefore, we can approximate for the number fraction
of self links $f_{s}\approx1/\left(1+f_{0}\sigma^{3/2}\right)$ and
for the fraction of crosslinks $f_{c}=1-f_{s}$. Note that from the
simulation data we obtain $f_{0}=26\pm1$.

Assuming no cross-correlations between cross- and self-linking, the
relative effective chain length $\beta(q)=\left\langle N_{x}(q)\right\rangle /N$
is computed from the contributions due to self-links \emph{and} crosslinks
\begin{equation}
\beta(q)=1-f_{s}\gamma q/N-f_{c}c{}^{2}\frac{\ln(2q+1)}{2}.\label{eq:beta(q)}
\end{equation}
The total average length reduction due to crosslinks and self-links
is computed from 
\begin{equation}
\frac{\left\langle N_{min}(q)\right\rangle }{N}=1-f_{s}\gamma q/N-f_{c}c\frac{\ln(2q+1)}{2}.\label{eq:Nmin}
\end{equation}
Both are used below to compare with the simulation data.

\subsection{Height reduction as a result of length reduction}

A simple analytical description of the height reduction of a polymer
brush upon crosslinking can be derived by assuming that the chain
length resisting the swelling of the brush is homogeneously reduced
to $\left\langle N_{x}(q)\right\rangle $, while the excluded volume
interactions that drive swelling depend on all monomers and thus remain
$\sim N$. This difference can be expressed in a Flory type free energy
per chain of form 
\begin{equation}
\Delta f\approx k_{B}T\left[\frac{3H^{2}}{2\left\langle N_{x}(q)\right\rangle b^{2}}+\mbox{v}N\left(\frac{N\sigma}{H}\right)\right].\label{Flory}
\end{equation}
Here, $\mbox{v}$ is the excluded volume parameter, $k_{B}$ the Boltzmann
constant, and $T$ the absolute temperature. The first term in the
square brackets describes the elastic contribution to free energy,
while the second term is the mean field estimate for pairwise monomer
contacts. Minimization with respect to brush height $H$ leads to
\begin{equation}
H\approx N^{2/3}\left(\mbox{v}\sigma\left\langle N_{x}(q)\right\rangle b^{2}\right)^{1/3}\approx H_{b}\beta(q)^{1/3},\label{height}
\end{equation}
which shows that the height reduction $H/H_{b}$ of the brush depends
solely on the average length reduction of the elastic chains in the
brush as given by $\beta(q)$.

\begin{figure}
\includegraphics[angle=270,width=0.8\columnwidth]{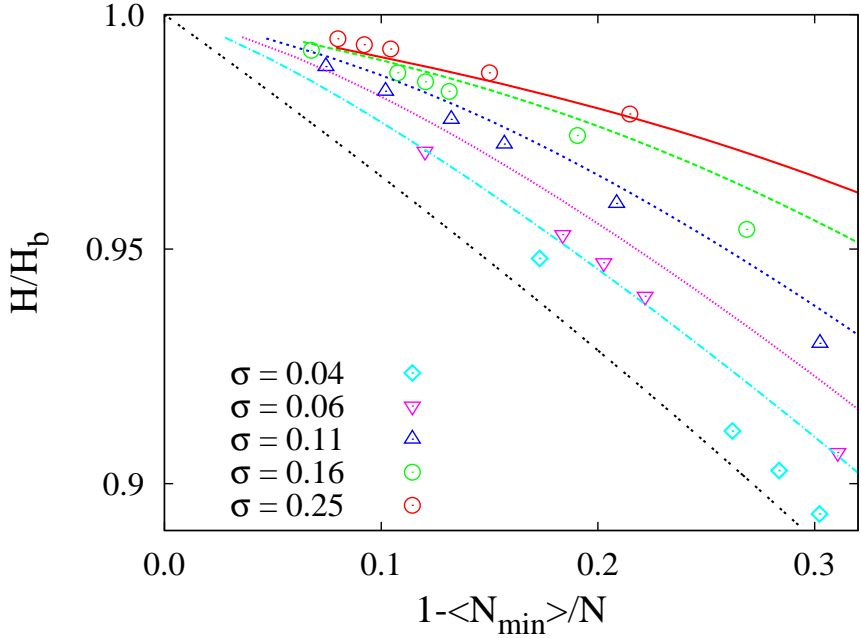} \caption{\label{fig:reduction4}Reduction of brush height upon crosslinking
as function of $\left\langle N_{min}(q)\right\rangle /N$ for brushes
with $N=64$ of different grafting densities. The color lines are
predictions based upon equation (\ref{height}) and (\ref{eq:Nmin})
as computed from the known number fractions $f_{c}$ and $f_{s}$
and the separately measured parameters $\gamma$ and $\left\langle |i-j|\right\rangle /N\approx c$
for cross- and self-links without additional fit parameter. The black
dotted line is the limit of $f_{c}=0$. }
\end{figure}

This prediction is tested against the simulation data in Figure \ref{fig:reduction4}.
The general trend of the height reduction of the brush is well predicted
by our mean field brush model. Note that the deviations between model
and data have the same trend as the overestimation of the length reduction
$\left\langle N_{min}(q)\right\rangle /N$ at large $q$ in Figure
\ref{fig:l_reduction}. This effect is apparently compensated at high
grafting density, since contributions of higher order virial terms
are neglected in equation (\ref{Flory}).

In order to highlight the main difference between our new model and
the uni-axial swelling of a network let us consider the case of crosslinking
monomers $i$ and $j$ with $i=j$. According to equation (\ref{eq:DeltaNeff}),
$\Delta N_{x}=0$. Thus, if there is $i=j$ for any reaction, the
height of the brush is expected to be constant \emph{independent}
of the degree of cross-linking, and therefore, independent of network
modulus. In consequence, a brush crosslinked in the swollen state
cannot be understood as a uni-axially swollen network. Note that the
general trend of the data in Figure (\ref{fig:reduction4}) clearly
supports our hypothesis: For increasing $\sigma,$ the samples contains
a larger number fraction of cross-links and thus, a larger modulus
at same length reduction. But the height reduction becomes smaller
for this larger modulus. Nevertheless, if the brush is linked in the
dry state for which the chains are collapsed or near to ideal conformations,
we still expect that these samples become equivalent to uni-axially
swollen networks upon swelling and that, therefore, the Flory-Rehner
model provides a reasonable approximation of experimental data.

The above results show that the relative change in the brush height
is rather minor when linking a brush in the swollen state (the cross-link
contribution is only a logarithmic correction). This is of particular
importance for experimental applications, since the lack of a collapse
of the brush upon crosslinking (no different scaling of the brush
height) allows to maintain a nearly unmodified function and interaction
with the environment, whereby the brush can be stabilized against
degradation.

Finally, it is interesting to observe that the frozen-in fluctuations
upon crosslinking a brush lead to a measurable height reduction. The
height reduction is achieved here by network imperfections and fluctuations
in the network structure. This is remarkable, since typically, network
defects are associated with a reduction of the network modulus. Similar
fluctuations and non-ideality are always present in any crosslinked
structure. Therefore, it will be worth investigating in a future work
to which extent non-ideal network structures can enhance the elastic
modulus of a polymer network or to which extent crosslinking of ordered
or stretched networks depends on monomer fluctuations.

\section{Summary\label{conclusion}}

In this work, we compare chain conformations of polymer brushes with
and without additional crosslinks between the chains. In order to
understand the height reduction of a brush upon crosslinking we discuss
the Flory-Rehner approach for the uni-axial swelling behavior of a
model network. If the grafting density is so high, that the number
$g$ of monomers per blob is smaller than the average length $n$
of a network strand, $g<n$, we expect no height reduction of the
brush as compared to the non-crosslinked height $H_{b}=N\sigma^{(1-\nu)/(2\nu)}b^{1/\nu}$.
For $g>n$ we expect that the brush shrinks with decreasing $n$ (for
increasing degree of crosslinking) according to $H\approx H_{b}\left(n/g\right)^{1/3}$.
We observe that it is difficult to reach the strongly crosslinked
regime when linking a brush in the swollen state, since in the swollen
state, a large portion of reaction connects monomers of the same chain
and thus, leads frequently to short dangling loops. Furthermore, the
simulation data does not collapse as function of $n/g$, which indicates
that at least an important correction to scaling is missing in the
Flory-Rehner approach.

An alternative model can be obtained by considering the particular
statistics of linking the same or neighboring chains. We find that
reactions between monomers of the same chain lead to short dangling
loops and a length reduction of the effective elastic chains in the
brush proportional to the degree of crosslinking $q$. Reactions between
different chains lead to a logarithmic correction for the effective
elastic chain length. Both corrections depend on local fluctuations
of the monomers in direction perpendicular to the grafting plane and
can be expressed in terms of a relative length $\beta(q)$ of the
effective elastic chain length with respect to $N$. This $\beta(q)$
can be incorporated in a Flory approach for the height of the brush
to yield $H(q)\approx H_{b}\beta(q)^{1/3}$ in good agreement with
the simulation data.

The results of the present work have impact on the understanding of
previous experimental results. For instance, our results suggest that
freezing a certain state of a switchable brush~\cite{Uhlmann:Brush2}
is only effective by linking both components together in a particular
switched state, since the collapse due to a selective crosslinking
in a co-solvent is rather minor. Furthermore, we expect a different
swelling behavior, if the chains are crosslinked in the dry state,
whereby the chain conformations are nearly ideal or collapsed. Finally,
our results indicate that cross-linking of ordered structures can
produce networks with a swelling equilibrium that is not controlled
by network modulus. This unexpected result is a challenge for theory
and probably a starting point to develop new tailor made materials.

\section{Acknowledgment}

M.L. acknowledges funding from the DFG under grant LA 2735/2-1.

\bibliographystyle{achemso} \addcontentsline{toc}{section}{\refname}\bibliographystyle{achemso}

\providecommand*{\mcitethebibliography}{\thebibliography} \csname
@ifundefined\endcsname{endmcitethebibliography} {\let\endmcitethebibliography\endthebibliography}{}

\end{document}